\documentclass[aps,prd,superscriptaddress,nofootinbib,amsmath,amsfonts,preprintnumbers,groupedaddress,showpacs,10pt,english]{revtex4-1}
\usepackage{amsmath}
\usepackage{amssymb}
\usepackage{babel}
\usepackage{wrapfig}
\usepackage{cancel}
\newcommand{\nn}{\nonumber \\}
\newcommand{\e}{\mathrm{e}}
\usepackage{relsize,exscale}
\makeatletter

\usepackage{array,multirow,graphicx}
\usepackage{dcolumn}
\usepackage{newlfont}
\usepackage{bm}
\usepackage[colorlinks,citecolor=blue,urlcolor=blue,linkcolor=blue]{hyperref}
\usepackage[figtopcap]{subfigure}
\usepackage{color}
\usepackage{verbatim}

\def\be{\begin{align}}
\def\ee{\end{align}}
\def\bea{\begin{eqnarray}}
\def\eea{\end{eqnarray}}
\def\bal{\begin{align}}
\def\eal{\end{align}}

\usepackage{scalerel}
\usepackage{tikz}
\usetikzlibrary{svg.path}
\definecolor{orcidlogocol}{HTML}{A6CE39}
\tikzset{
 orcidlogo/.pic={
 \fill[orcidlogocol] svg{M256,128c0,70.7-57.3,128-128,128C57.3,256,0,198.7,0,128C0,57.3,57.3,0,128,0C198.7,0,256,57.3,256,128z};
 \fill[white] svg{M86.3,186.2H70.9V79.1h15.4v48.4V186.2z}
 svg{M108.9,79.1h41.6c39.6,0,57,28.3,57,53.6c0,27.5-21.5,53.6-56.8,53.6h-41.8V79.1z M124.3,172.4h24.5c34.9,0,42.9-26.5,42.9-39.7c0-21.5-13.7-39.7-43.7-39.7h-23.7V172.4z}
 svg{M88.7,56.8c0,5.5-4.5,10.1-10.1,10.1c-5.6,0-10.1-4.6-10.1-10.1c0-5.6,4.5-10.1,10.1-10.1C84.2,46.7,88.7,51.3,88.7,56.8z};}}
\newcommand\orcid[1]{\href{https://orcid.org/#1}{\mbox{\scalerel*{
\begin{tikzpicture}[yscale=-1,transform shape]
\pic{orcidlogo};
\end{tikzpicture}
}{|}}}}
\graphicspath{{./}{Figs/}}
\begin{document}
\date{\today}
%%%%%%%%%%%%%%%%%%%%%%%%%%%%%%%%%%%%%%%%%%%%%%%%%%%%%%%%%%%%%%%%%%%%%%%%%%%%%%%%%%%%%%%%%%%%%%%%%%%%%%%
\title{Wormhole  solution free of ghosts in Einstein's  gravity with two scalar fields}
\author{Shin'ichi~Nojiri~\orcid{0000-0002-0773-8011}}
\email{nojiri@gravity.phys.nagoya-u.ac.jp}
\affiliation{Department of Physics, Nagoya University, Nagoya 464-8602,
Japan \\
\& \\
Kobayashi-Maskawa Institute for the Origin of Particles and the Universe,
Nagoya University, Nagoya 464-8602, Japan }
\author{G.~G.~L.~Nashed~\orcid{0000-0001-5544-1119}}
\email{nashed@bue.edu.eg}
\affiliation {Centre for Theoretical Physics, The British University in Egypt, P.O. Box
43, El Sherouk City, Cairo 11837, Egypt}
%%%%%%%%%%%%%%%%%%%%%%%%%%%%%%%%%%%%%%%%%%%%%%%%%%%%%

\begin{abstract}
In this paper, we construct models that admit the traversable wormhole geometries in the framework of Einstein's gravity with two scalar fields.
As well known, the energy conditions are broken and we show that there appears a ghost.
The ghost can be, however, eliminated by imposing a constraint on the ghost field, which is a scalar.
The constraint is similar to the mimetic one proposed by Chamseddine and Mukhanov to construct an alternative description of cold dark matter.
We explicitly show that there does not appear any unstable mode although the energy conditions are broken.
Therefore we obtain a model that realizes the traversable and stable wormhole.
\end{abstract}
\maketitle
%\keywords{ Consistent mimetic gravitational theory; regular black holes; flat rotation curve, thermodynamics.}
%\pacs{ 04.50.Kd, 98.80.-k, 04.80.Cc, 95.10.Ce, 96.30.-t}
%\end{abstract}
%\vspace{0.5cm} \hrule
%\def\thefootnote{\arabic{footnote}}
%\setcounter{footnote}{0}
%\maketitle
%%%%%%%%%%%%%%%%%%%%%%%%%%%%%%%%%%%%%%%%%%%%%%%%%%%%%

%\newpage
\section{Introduction}\label{Sec1}

Recently, observations involving the relativistic collision of two compact objects have resulted in the production of gravitational waves (GWs). These waves serve as invaluable instruments for examining the properties of the colliding entities.
Moreover, the recent findings reported by LIGO~\cite{LIGOScientific:2016aoc, LIGOScientific:2016sjg, LIGOScientific:2017bnn, LIGOScientific:2017ycc, LIGOScientific:2017vwq} have provided compelling evidence that the field of gravitational wave astronomy will play a substantial role in advancing our understanding of gravitational interactions and extreme-gravity astrophysical phenomena. However, despite these notable advancements, recent observations have not ventured into the intricate intricacies of spacetime beyond the photon sphere.
The expected forthcoming gravitational wave (GW) observations are in a position to record the ringdown phase. This phase is recognized by a series of damped oscillatory modes in the initial stages, commonly known as quasinormal modes (QNMs)~\cite{Vishveshwara:1970zz, Kokkotas:1999bd, Berti:2009kk, Mazur:2001fv, Morris:1988tu, Damour:2007ap, Konoplya:2011qq}.
This stage has the capacity to provide vital information about the composition of compact objects~\cite{Cardoso:2016rao}, particularly elucidating the physics in the vicinity of black holes' event horizons (BHs) and the potential presence of unexpected structural characteristics.
Future gravitational wave (GW) observations hold the promise of providing us with a deeper understanding of compact objects that differ from black holes (BHs). These distinct compact entities, which lack event horizons, are commonly known as exotic compact objects (ECOs)~\cite{Mazur:2001fv, Damour:2007ap, Holdom:2016nek}.

Among the notable exotic compact object (ECO) solutions, wormholes (WHs) stand out. They are solutions to the Einstein equations that enable connections between different regions of the Universe or even between entirely separate Universes~\cite{Misner:1960zz, Wheeler:1957mu}.
Although they have distinct causal structures when compared to black holes (BHs), wormholes (WHs) { can} possess photon spheres.
As a result, in gravitational wave (GW) data, the early stage of the ringdown signal has the potential to mask the ability to distinguish between wormholes (WHs) and black holes (BHs).
Previous research has delved into the examination of Lorentzian wormholes within the framework of General Relativity (GR), as documented in prior studies~\cite{Bronnikov:1973fh, Ellis:1973yv, Azad:2022qqn,Morris:1988tu, Sahni:1999gb, Carroll:2000fy, Peebles:2002gy,Nashed:2011fg, Gonzalez-Diaz:2002xox, Gonzalez:2022ote, Nashed:2010ocg, Poisson:1995sv}.
In these investigations, the establishment of conditions for traversable wormholes was accomplished by introducing a static spherically symmetric metric.
Significantly, these conditions necessitate the inclusion of exotic or phantom matter, which violates the null energy condition (NEC).
Wormholes that incorporate ordinary matter adhering to the null energy condition (NEC)\cite{Poisson:1995sv, Antoniou:2019awm, Mehdizadeh:2015dta} have been introduced and explored within the framework of modified gravity theories. These theories encompass Brans-Dicke theory\cite{Nandi:1997mx, Blazquez-Salcedo:2018ipc,Lobo:2010sb, Sushkov:2011zh, Papantonopoulos:2019ugr}, $f(R)$ gravity~\cite{Garcia:2010xb, MontelongoGarcia:2010xd}, Einstein-Gauss-Bonnet theory~\cite{Mehdizadeh:2015jra, KordZangeneh:2015dks}, Einstein-Cartan theory, and general scalar-tensor theories~\cite{Bronnikov:2015pha, Bronnikov:2016xvj, Mehdizadeh:2017dhb, Yaqin:2017bij}.
It's worth highlighting that wormhole (WH) solutions within the context of $f(R)$ theories were extensively investigated in the work presented in \cite{Karakasis:2021tqx}.
Furthermore, wormholes (WHs) with a self-interacting scalar field were the focus of scrutiny in the study presented in \cite{Anabalon:2012tu}.

Another pivotal element concerning a WH involves the breach of energy conditions within the context of GR,
at least in the vicinity of the WH's throat \cite{Morris:1988cz, Lobo:2007zb}.
This implies that it is necessary to possess a quantity of exotic matter (where the stress-energy tensor (SET) of matter contravenes the null energy condition (NEC))
in order to maintain the stability of the WH throat.
As a result, in certain reference frames, the energy density of matter can be perceived as negative. It is evident that there has been significant interest
in constructing WHs while minimizing the requirement for exotic matter \cite{Visser:2003yf}.
In reference~\cite{Visser:2003yf}, the authors have theoretically demonstrated that it is possible to minimize the exotic matter requirement to an infinitesimal level
and confine it precisely at the throat of the WH by carefully selecting the WH's geometry.
The mathematical procedure is referred to as the ``cut and paste technique'', and the resulting WH is termed a ``thin-shell WH''.
Research on thin-shell WHs can be located in references \cite{Visser:1989kh, Visser:1989kg, Lobo:2003xd, Dias:2010uh}.
Nandi et al.~\cite{Nandi:2004ku} subsequently introduced an enhanced quantification method to precisely determine the amount of exotic matter present in a specific spacetime.
Consequently, numerous arguments have been presented to substantiate the breach of energy conditions.
In this context, the use of the phantom energy equation of state (EoS) has been employed to maintain traversable WHs,
as demonstrated in references \cite{Konoplya:2021hsm,Sushkov:2005kj, Lobo:2005us, Gonzalez:2009cy}. An interpretation of
half-wormholes in the bulk with gauge field { was also investigated in} \cite{Choudhury:2021nal}

One intriguing aspect to contemplate is that within the phantom regime, the energy density grows over time, offering a conceptual basis for the existence of WHs.
The stability analysis of a phantom WH geometry has been explored in reference~\cite{Lobo:2005yv}.
While an array of WH solutions has been identified for the generalized Chaplygin gas \cite{Lobo:2005vc, Kuhfittig:2009mx}, changing cosmological constants \cite{Rahaman:2006xa},
polytropic phantom energy \cite{Jamil:2010ziq}, and ghost scalar fields \cite{Carvente:2019gkd}.
Subsequently, the phantom energy EoS was employed to formulate precise, evolving WH structures in reference \cite{Cataldo:2008pm}.
In this theoretical framework, it was determined that phantom energy can facilitate the existence of evolving WHs.

Nonetheless, physicists consistently strive to evade energy condition violations or provide appropriate justifications for them.
However, up to the present time, constructing a static WH geometry that complies with the energy conditions remains an unattained goal.
Consequently, scientists are exploring various strategies to address this challenge.
This observation prompted consideration of the potential existence of WH solutions within alternative theories of gravity.
Examples include higher-order gravity theories~\cite{Hochberg:1990is,Ghoroku:1992tz}, cosmological WHs in higher dimensions~\cite{Zangeneh:2014noa},
and the Einstein-Gauss-Bonnet theory~\cite{Cuyubamba:2018jdl,Bhawal:1992sz,Maeda:2008nz,Mehdizadeh:2015jra}.
When examining $f(R)$ gravity, it is conceivable to theoretically create traversable WHs without the necessity of exotic matter \cite{Pavlovic:2014gba, Lobo:2009ip}
or by utilizing dark matter as a source \cite{Muniz:2022eex}.
Alternatively, researchers can explore the quest for WHs within the framework of third-order Lovelock gravity \cite{KordZangeneh:2015dks, Mehdizadeh:2016nna},
hybrid metric-Palatini gravity~\cite{Rosa:2021yym, KordZangeneh:2020ixt}, $f(Q)$ gravity~\cite{Banerjee:2021mqk, Parsaei:2022wnu, Hassan:2022hcb},
and extended theories of gravity \cite{DeFalco:2021klh, DeFalco:2021ksd}.
The investigation of traversable WHs within the realm of $f(R, T)$ gravity was conducted in references~\cite{Moraes:2017mir, Elizalde:2018frj, Moraes:2019pao}.
Simultaneously, in references \cite{Rosa:2022osy, Banerjee:2020uyi}, authors identified precise WH solutions in $f(R, T)$ gravity without the need for exotic matter.
It is the aim of the present study to derive WH in the frame of Einstein's gravity coupled with two scalar fields.

In \cite{Nojiri:2020blr}, a general formulation to construct a model that admits arbitrarily given spherically symmetric and time-dependent geometry as a solution has been given
in the framework of Einstein's gravity coupled with two scalar fields.
We apply the formulation to the static wormhole formulation.
As expected, the model includes ghost mode.
The ghost mode often plays the role of the phantom and is consistent with the breakdown of the energy conditions.
The ghost mode has, however, negative kinetic energy classically and generates negative norm states as a quantum theory.
Therefore the existence of the ghost mode tells that the model is physically inconsistent.
In this paper, we eliminate such ghosts using the mimetic constraint and make the ghost mode non-dynamical.
We show that there does not appear unstable mode corresponding to the ghost mode although the energy conditions are still broken.
This may tell that the breakdown of the energy conditions could not always imply physical inconsistency.

The organization of this paper is as follows:
In the next section, based on the formulation in \cite{Nojiri:2020blr}, we construct a model whose solutions include a well-known wormhole geometry.
We show that there appears a ghost in the model.
In Section~\ref{Sec3}, we show that one of the two scalar fields can be canonical and not a ghost.
Because another one is a ghost in general, we propose a model to make the scalar field non-dynamical by imposing the mimetic constraint, and as a result,
unstable modes corresponding to the existence of the ghost disappear.
In Section~\ref{Sec4}, we confirm the absence of the ghost mode by using the perturbation from the wormhole geometry although the energy conditions are broken.
The breakdown of the energy conditions may not always imply any physical inconsistency.
The last section is devoted to the summary and discussion.

\section{Wormhole based on Einstein's theory with two scalar fields}\label{Sec2}

Einstein's GR with two scalar fields $\phi$ and $\chi$ is described by the action as follows~\cite{Nojiri:2020blr}
%\footnote{Throughout this study, we will use the relativistic units in which we will put $\kappa=1$.},
\begin{align}
\label{I8}
S_{\mathrm{GR} \phi\chi} = \int d^4 x \sqrt{-g} & \left[ \frac{R}{2\kappa^2}
 - \frac{1}{2} \, A (\phi,\chi) \partial_\mu \phi \partial^\mu \phi
 - B (\phi,\chi) \, \partial_\mu \phi \partial^\mu \chi \right. \nn %&\nn
& \left. \quad - \frac{1}{2} \, C (\phi,\chi) \partial_\mu \chi \partial^\mu \chi - V (\phi,\chi)\right] \, .
\end{align}
In this context, $g$ represents the determinant of the metric tensor $g_{\mu\nu}$, $R$ denotes the Ricci scalar, and $V(\phi, \chi)$ represents the potential of the scalar doublet. The values of the coefficients $A$, $B$, and $C$ are contingent upon the properties of the scalars.
Upon varying the action (\ref{I8}) with respect to the metric $g_{\mu\nu}$, we derive the ensuing Einstein equation:
\begin{align}
\label{I9}
\frac{1}{\kappa^2} \left( R_{\mu\nu} - \frac{1}{2} g_{\mu\nu} R \right) = &\, A (\phi,\chi) \partial_\mu \phi \partial_\nu \phi
+ B (\phi,\chi) \left( \partial_\mu \phi \partial_\nu \chi + \partial_\nu \phi \partial_\mu \chi \right)
+ C (\phi,\chi) \partial_\mu \chi \partial_\nu \chi\nn
&\, -g_{\mu\nu} \left[
\frac{1}{2}\, A (\phi,\chi) \partial_\rho \phi \partial^\rho \phi
+ B (\phi,\chi) \partial_\rho \phi \partial^\rho \chi
 + \frac{1}{2} \, C (\phi,\chi) \partial_\rho \chi \partial^\rho \chi + V (\phi,\chi)\right] \, ,
\end{align}
Through the variation of action (\ref{I8}) concerning the scalar fields $\phi$ and $\chi$, we acquire the following expressions:
\begin{align}
\label{I10}
0 =&\, \frac{A_\phi}{2}\, \partial_\mu \phi \partial^\mu \phi
+ A \nabla^\mu \partial_\mu \phi + A_\chi \partial_\mu \phi \partial^\mu \chi
+ \left( B_\chi - \frac{1}{2} \, C_\phi \right)\partial_\mu \chi \partial^\mu \chi + B \nabla^\mu \partial_\mu \chi - V_\phi \, ,\\
\label{I10b}
0 =&\, \left( - \frac{1}{2} \, A_\chi + B_\phi \right)
\partial_\mu \phi \partial^\mu \phi + B \nabla^\mu \partial_\mu \phi + \frac{1}{2} \, C_\chi \partial_\mu \chi \partial^\mu \chi
+ C \nabla^\mu \partial_\mu \chi + C_\phi \partial_\mu \phi \partial^\mu \chi - V_\chi\, .
\end{align}
Here, let us define $A_\phi$ as $\partial A(\phi,\chi)/\partial \phi$, and similarly for other derivatives. It's worth noting that Eqs.(\ref{I10}) and (\ref{I10b}) can be derived from the Bianchi identity in conjunction with Eq.(\ref{I9}).

In the following, we identify
\begin{align}
\label{TSBH1}
\phi=t\, , \quad \chi=r\, .
\end{align}
As explained in the reference \cite{Nojiri:2020blr}, making the assumption (\ref{TSBH1}) doesn't result in any loss of generality.
In the case of a spacetime with a general spherically symmetric yet time-dependent solution, the scalar fields $\phi$ and $\chi$ exhibit dependencies on both the time coordinate, denoted as $t$, and the radial coordinate, denoted as $r$.
In the context of a given solution, the specific dependencies of $\phi$ and $\chi$ on both the time variable $t$ and the radial variable $r$ are determined as functions: $\phi = \phi(t, r)$ and $\chi = \chi(t, r)$.
We may redefine the scalar fields to replace $t$ and $r$ with new scalar fields, $\tilde\phi$ and $\tilde\chi$,
$\phi\left( \tilde\phi, \tilde\chi \right) \equiv \phi\left( t=\tilde\phi, r=\tilde\chi \right)$ and
$\chi\left( \tilde\phi, \tilde\chi \right) \equiv \chi\left( t=\tilde\phi, r=\tilde\chi \right)$.
Subsequently, we can associate the new scalar fields with the time and radial coordinates in (\ref{TSBH1}). The transformation of variables from $\left(\phi,\chi\right)$ to $\left(\tilde\phi,\tilde\chi\right)$ can be integrated into the redefinitions of $A$, $B$, $C$, and $V$ within the action (\ref{I8}). This demonstrates that making the assumption (\ref{TSBH1}) doesn't lead to any loss of generality. Furthermore, as we will observe later, even when $\phi$ is identified with $t$ a static spacetime can still be achieved.

Now we consider the following spherically symmetric line-element
\begin{align}
\label{1}
ds^2=-\e^{2\Phi(r)}dt^2+ \left( 1- \frac{b(r)}{r} \right)^{-1} dr^2+r^2 \left( d\theta^2+\sin^2\theta d\phi^2 \right)\,.
\end{align}
Here, $\Phi\left(r\right)$ and $b\left(r\right)$ represent arbitrary functions of the radial coordinate, and they are referred to as the redshift and shape functions, respectively.
At the minimum wormhole throat, denoted as $r_0$ with the condition $b(r_0) = r_0$, the wormhole serves as a connection between two distinct Universes,
and the radial coordinate range follows the inequality $0 < r_0 \leq r \leq \infty$.
To prevent the formation of an event horizon or any singularities at the wormhole throat $r_0$, the redshift function should remain well-defined across all points.
To ensure the traversability of the wormhole, two conditions must be met:
{ $b\left(r\right)- rb'\left(r\right)> 0$, and $1-\frac{b\left(r\right)}{r}> 0$
(the second condition is required except the throat at $ r = r_0$)}.
(where $'$ represents the derivative with respect to $r$).
Additionally, for asymptotically flat wormhole solutions, the condition $\frac{b\left(r\right)}{r}\to 0$ as $r \to \infty$ is imposed, as outlined in \cite{Shamir:2017byl}.

%%%%%%%%%%%%%%%%%%%%%%%%%

Applying the field equation (\ref{I8}) to the line element (\ref{1}), we obtain,
\begin{align}
\label{fe}
\frac{b' \left( r \right) }{ r^2} =&\, \frac{ \left( \left( \left( r-b \left( r \right) \right) C \left(r\right) +2V \left(r\right) r \right) \e^{2\Phi \left( r \right)}
+A \left(r\right) r \right) \kappa^2}{2 \e^{2\Phi \left( r \right)} r} \,,\nonumber\\
B \left(r\right)=&\, 0\,,\nonumber\\
\frac{2 r \left( r-b \left( r \right) \right) \Phi' \left( r \right) -b \left( r \right) }{r^3}
=&\, \frac{ \left( \left( \left( r-b \left( r \right) \right) C \left( r \right) -2 V \left(r\right) r \right) \e^{2\Phi \left( r \right)}
+A \left(r\right) r \right) \kappa^2}{2 \e^{2\Phi \left( r \right)}r} \,,\nonumber\\
\frac{1}{r^3} \Bigl\{ 2 r^2 \left( r-b \left( r \right) \right)\Phi'' \left( r \right) & +2 \Bigl( r \left( r-b \left(r \right) \right)\Phi' \left( r \right)
 - \frac{b' \left( r \right)r}{2} + \frac{b \left( r \right)}{2} \Bigr) \left( 1+ \Phi' \left( r \right) r \right) \Bigr\} \nonumber\\
=&\, {\frac{ \left( \left( \left( -r + b \left( r \right) \right) C \left(r\right) -2\,V \left( r \right) r \right) \e^{2\Phi \left( r \right) } +A \left(r\right) r \right) \kappa^2}
{\e^{2\Phi \left( r \right) } r}} \, .
\end{align}
Assuming the WH geometry to has the form \cite{Tangphati:2020mir, Capozziello:2020zbx}
\begin{align}
\label{WH}
\Phi\left(r\right)=\frac{r_0}{2r}\, , \quad b\left(r\right)=r\e^{-\gamma \left( r-r_0 \right)}\, ,
\end{align}
with positive constants $r_0$ and $\gamma$ and substituting these expressions into Eq.~(\ref{fe}), we obtain
\begin{align}
\label{fes}
A \left(r\right) =&\, \frac{\e^{\frac{r_0}{r}}}{4 \kappa^2 r^4 \e^{\gamma r} } \left( -2 r_0 r \e^{\gamma r_0} + 4 r^2\e^{\gamma r_0} + 2 r_0 r\e^{\gamma r}
 -2 \gamma r^3 \e^{\gamma r_0} - {r_0}^2 \e^{\gamma r_0} - \gamma r_0 r^2 \e^{\gamma r_0} +{r_0}^2\e^{\gamma r} \right)\,,\nonumber\\
B \left(r\right)=&\, 0\,,\nonumber\\
C \left(r\right) =&\, - \frac{4 r^2\e^{\gamma r_0}-6 r_0 r \e^{\gamma r_0} + 6 r_0 r\e^{\gamma r} + 2 \gamma r^3 \e^{\gamma r_0} - {r_0}^2 \e^{\gamma r_0}
 - \gamma r_0 r^2 \e^{\gamma r_0} +{r_0}^2 \e^{\gamma r}}{4\kappa^2 r^4 \left( \e^{\gamma r}-\e^{\gamma r_0} \right) }\,,\nonumber\\
V \left(r\right) =&\, \frac{-r_0\e^{\gamma r_0}+2 r \e^{\gamma r_0} + r_0\e^{\gamma r} - \gamma r^2 \e^{\gamma r_0}}{{2\kappa}^2 r^3\e^{\gamma r}}\,.
\end{align}
We are assuming $r\geq r_0$.
Eq.~(\ref{fes}) tells that if we consider the model,
\begin{align}
\label{feschi}
A \left(\chi\right) =&\, \frac{\e^{\frac{r_0}{\chi}}}{4 \kappa^2 \chi^4 \e^{\gamma \chi} } \left( -2 r_0 \chi \e^{\gamma r_0} + 4 \chi^2\e^{\gamma r_0} + 2 r_0 \chi\e^{\gamma \chi}
 -2 \gamma \chi^3 \e^{\gamma r_0} - {r_0}^2 \e^{\gamma r_0} - \gamma r_0 \chi^2 \e^{\gamma r_0} +{r_0}^2\e^{\gamma \chi} \right) \,,\nonumber\\
B \left(\chi\right)=&\, 0\,,\nonumber\\
C \left(\chi\right) =&\, - \frac{4 \chi^2\e^{\gamma r_0}-6 r_0 \chi \e^{\gamma r_0} + 6 r_0 \chi\e^{\gamma \chi} + 2 \gamma \chi^3 \e^{\gamma r_0} - {r_0}^2 \e^{\gamma r_0}
 - \gamma r_0 \chi^2 \e^{\gamma r_0} +{r_0}^2 \e^{\gamma \chi}}{4\kappa^2 \chi^4 \left( \e^{\gamma \chi}-\e^{\gamma r_0} \right) } \,,\nonumber\\
V \left(\chi\right) =&\, \frac{-r_0\e^{\gamma r_0}+2 r \e^{\gamma r_0} + r_0\e^{\gamma \chi} - \gamma \chi^2 \e^{\gamma r_0}}{{2\kappa}^2 \chi^3\e^{\gamma \chi}} \,,
\end{align}
the wormhole spacetime given by (\ref{1}) with (\ref{WH}) and (\ref{TSBH1}) is an exact solution of the model.

\begin{figure*}
\centering
\subfigure[~A$\&$ C]{\label{fig:A}\includegraphics[scale=0.3]{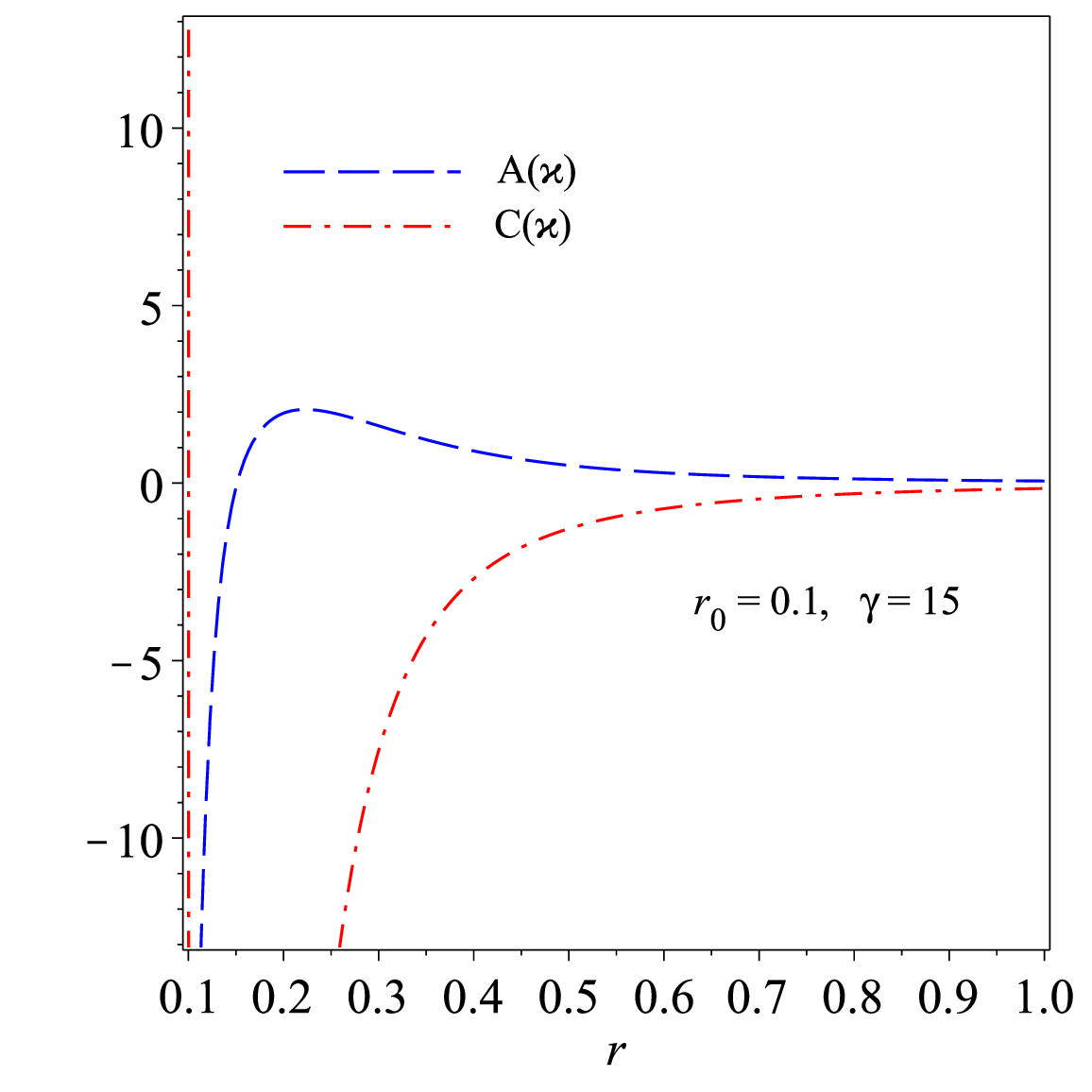}}\hspace{0.5cm}
\subfigure[~Density, radial and tangential pressures]{\label{fig:rpp}\includegraphics[scale=0.3]{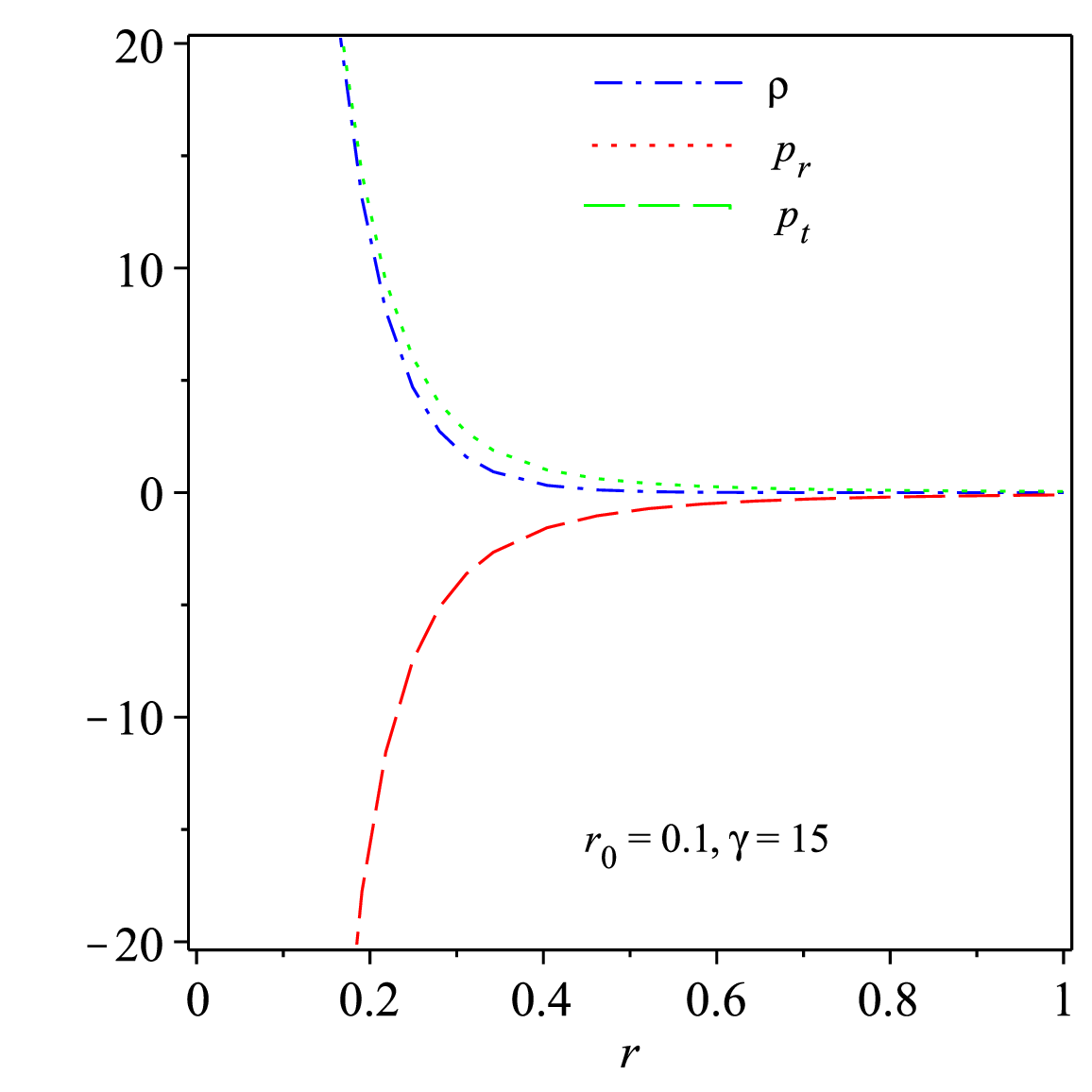}}\\
\subfigure[~some components of the energy conditions]{\label{fig:EC}\includegraphics[scale=0.3]{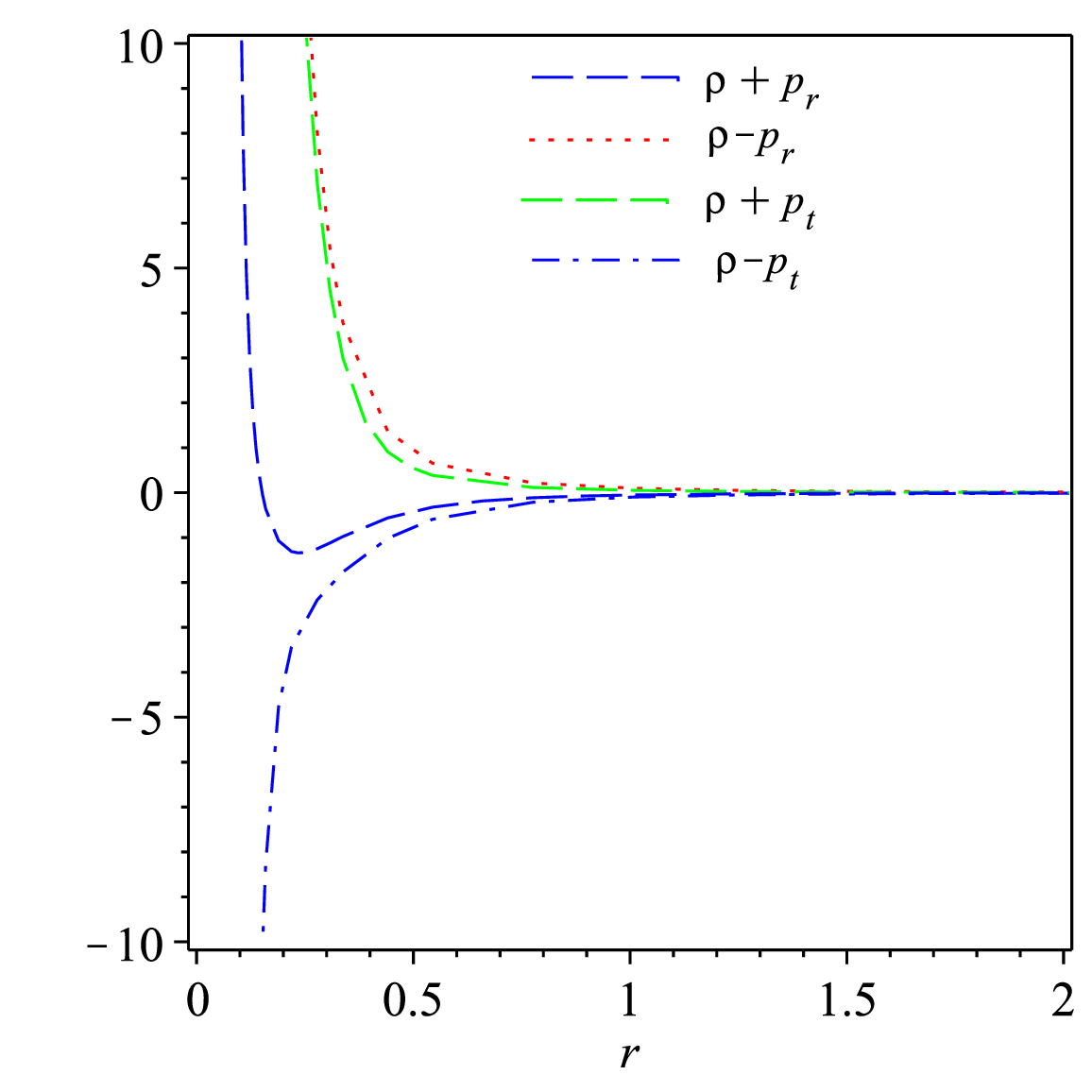}}\\
\caption{\subref{fig:A} The general behaviors of the two functions $A$ and $C$ as $r\to r_0$;
\subref{fig:rpp} is the behavior of the density, radial, and tangential pressure;
\subref{fig:EC} represents some components of the energy conditions.}
\label{Fig:dens_press}
\end{figure*}

We should note both $A\left(r\right)$ and $C\left(r\right)$ become negative in some region of $r$, as shown in Fig.~\ref{Fig:dens_press} \subref{fig:A}, for example, when $r\to r_0$, we obtain,
\begin{align}
\label{fes2}
A \left(r\right) \to \frac{\e^{\frac{r_0}{r}}}{4 \kappa^2 {r_0}^2} \left( 4 - 3\gamma r_0 \right)\,, \quad
C \left(r\right) \to -\frac{4 + \gamma r_0 } {4\kappa^2 \gamma {r_0}^2 \left( r - r_0 \right) }\, .
\end{align}
Therefore $C\left(r\right)$ is negative and $A\left(r\right)$ becomes negative if $4 - 3\gamma r_0<0$.

When $A\left(r\right)$ or $C\left(r\right)$ is negative, the scalar field $\phi$ or $\chi$ becomes a ghost, which generates the breakdown of the energy conditions in general.
Using Eq.~(\ref{feschi}), we obtain the form of the energy density $\rho$ and the radial and tangential components of the pressure, $p_\mathrm{r}$ and $p_\mathrm{t}$ as follows, \footnote{
We may express Eq.~(\ref{I9}) as $G_{\mu\nu}=\kappa^2 T^\mathrm{sc}_{\mu\nu}$ by using the Einstein tensor $G_{\mu\nu}:={ R}_{\mu\nu}-g_{\mu\nu}{R}/2$ and the energy-momentum tensor
of the scalar field $T^\mathrm{sc}_{\mu\nu}$.
By writing the energy-momentum tensor as $T^{\mathrm{sc} \nu}_{\ \ \mu}=\mathrm{diag} \left(-\rho, p_\mathrm{r}, p_\mathrm{t},p_\mathrm{t} \right)$,
we can extend the energy conditions to Einstein's gravity with scalar two fields as follows,
\begin{itemize}
 \item[1.] $ {\rho} + {p}_r > 0$, ${\rho} +{p}_t > 0$, and ${\rho}\geq 0$, which is the Weak Energy Condition (WEC).
 \item[2.] ${\rho}+ {p}_t \geq 0$, and ${\rho}+ {p}_r \geq 0$, which is the Null Energy Condition (NEC).
 \item[3.] ${\rho}+{p}_r \geq 0$, ${\rho}+{p}_t \geq 0$, and ${\rho} +{p}_r+2{p}_t \geq 0$, which is the Strong Energy Condition (SEC).
 \item[4.] ${\rho}-{p}_r \geq 0$, ${\rho} -{p}_t \geq 0$,and ${\rho}\geq 0$, which is the Dominant Energy Conditions (DEC).\\
\end{itemize}
}
\begin{align}
\label{comp}
\rho=&\, \frac{- \e^{\gamma r_0} + \e^{\gamma r_0} \gamma r + \e^{-\gamma \left( r-2 r_0 \right)}
 - \e^{-\gamma \left( r-2 r_0 \right)}r \gamma}{r^2 \left( \e^{\gamma r} - \e^{\gamma r_0} \right) } \,,\nonumber\\
p_\mathrm{r}=&\, - \frac{r \e^{\gamma r_0} + r_0 \e^{\gamma r} - 2 r_0 \e^{\gamma r_0} - \e^{-\gamma \left( r -2 r_0 \right)} r
+ \e^{-\gamma \left( r-2 r_0 \right)} r_0}{r^3 \left( \e^{\gamma r} - \e^{ \gamma r_0} \right) } \,,\nonumber\\
p_\mathrm{t}=&\, \frac{ \left( {r_0}^2+ \left( \gamma r^2 + 2r \right) r_0-2\gamma r^3 \right) \e^{-\gamma \left( r-2r_0 \right)}
+ \left(2\gamma r^3 -2{r_0}^2- \left( \gamma r^2+4r \right) r_0 \right) \e^{\gamma r_0} + r_0 \left( 2r+r_0 \right) \e^{\gamma r}}{4r^4 \left( \e^{\gamma r} - \e^{\gamma r_0} \right) }\,.
\end{align}
Then we find the energy conditions are broken when $r\gtrsim r_0$ as shown in Fig.~\ref{Fig:dens_press} \subref{fig:rpp} and Fig.~\ref{Fig:dens_press} \subref{fig:EC}.

\section{Eliminating ghosts}\label{Sec3}

We consider the conditions that $A\left(r\right)$ becomes non-negative.
We define a function $N\left(r\right)$ by $A\left(r\right)=\frac{\e^{\frac{r_0}{r}}N\left(r\right)}{4 \kappa^2 r^4 \e^{\gamma r} } $, that is
\begin{align}
\label{N1}
N\left(r\right)\equiv&\, -2 r_0 r \e^{\gamma r_0} + 4 r^2\e^{\gamma r_0} + 2 r_0 r\e^{\gamma r}
 -2 \gamma r^3 \e^{\gamma r_0} - {r_0}^2 \e^{\gamma r_0} - \gamma r_0 r^2 \e^{\gamma r_0} +{r_0}^2\e^{\gamma r} \, .
\end{align}
Then we find
\begin{align}
\label{N1B}
N''''\left(r\right) =&\, 8 \gamma^3 r_0 \e^{\gamma r} + 2 \gamma^4 r_0 r\e^{\gamma r}
+ \gamma^4 {r_0}^2\e^{\gamma r} > 0 \, ,
\end{align}
and
\begin{align}
\label{N2}
N(r_0)= &\, \left(4 - 3 \gamma r_0 \right) {r_0}^2\e^{\gamma r_0} \, , \quad
N'(r_0) = \left( 8 - 5 \gamma r_0 \right) r_0 \e^{\gamma r_0} \, ,\quad
N''(r_0) = \left( 8 - 10 \gamma r_0 + 3 \gamma^2 {r_0}^2 \right) \e^{\gamma r_0} \, , \nonumber \\
N'''(r_0) =&\, \gamma \left( -12 + 6 \gamma r_0 + 3 \gamma^2 {r_0}^2 \right) \e^{\gamma r_0} \, ,
\end{align}
The conditions $N(r_0)\geq 0$, $N'(r_0)\geq 0$, $N''(r_0)\geq 0$, and $N'''(r_0)\geq 0$ give
$\gamma r_0 \leq \frac{4}{3}\approx1.33$, $\gamma r_0 \leq \frac{8}{5}=1.6$, $\gamma r_0 \leq \frac{4}{3}$ or $\gamma r_0 \geq 2$, and
$\gamma r_0 \leq - 1 - \sqrt{5}$ or $\gamma r_0 \geq - 1 + \sqrt{5}\approx 1.236$, respectively.
Therefore if
\begin{align}
\label{N3}
\frac{4}{3} \geq \gamma r_0 \geq - 1 + \sqrt{5}\, ,
\end{align}
$N\left(r\right)$ and therefore $A\left(r\right)$ are not negative, which tells $\phi$ is not ghost but canonical scalar field.
For the WH to be traversable, $r_0$ should be large enough.
Such $r_0$ can be realized by choosing $\gamma$ small enough to satisfy the condition (\ref{N3}).

On the other hand, $\chi$ is a ghost in general.
In order to avoid the ghost, we propose a model where the scalar field $\chi$ becomes non-dynamical by imposing the mimetic constraint on $\chi$
as follows\footnote{An alternative conceptualization to the concept of cold dark matter emerges through the mimetic modification of General Relativity (GR),
as originally introduced by Chamseddine and Mukhanov~\cite{Chamseddine:2013kea}.
Subsequent investigations into this theoretical framework have been conducted in a series of
works~\cite{Chamseddine:2014vna, Nojiri:2022cah, Myrzakulov:2015kda, Nashed:2021hgn, Nashed:2023jdf, Nashed:2023fzp, Myrzakulov:2015qaa, Vagnozzi:2017ilo,
Nashed:2021ctg, Casalino:2018tcd,Casalino:2018wnc,Sebastiani:2016ras,Nashed:2021pkc}.
In their paper \cite{Chamseddine:2013kea}, Chamseddine and Mukhanov isolated the conformal degree of freedom of Einstein-Hilbert gravity in a covariant way,
and in the resulting theory, the physical metric is defined with the account of an auxiliary scalar field, which appears through its first derivatives.
{ In this sense, the addition of the term (\ref{Smim}) to the action (\ref{I8}) may be regarded with a modification of Einstein'is gravity. }},
\begin{align}
\label{mimeticchi}
\left( 1 - \frac{b(\chi)}{\chi} \right) g^{\mu\nu} \partial_\mu \chi \partial_\nu \chi = 1\, ,
\end{align}
whose solution is consistently $\chi=r$.
The constraint can be realized by introducing a multiplier field $\lambda$ and add the following term $S_\mathrm{mim}$ to the action (\ref{I8}),
$S_{\mathrm{GR} \phi\chi} \to S_{\mathrm{GR} \phi\chi} + S_\mathrm{mim}$,
\begin{align}
\label{Smim}
S_\mathrm{mim} \equiv \int d^4 x \sqrt{-g} \lambda \left\{ \left( 1 - \frac{b(\chi)}{\chi} \right) \partial_\rho \chi \partial^\rho \chi - 1 \right\} \, .
\end{align}
By the term (\ref{Smim}), Eq.~(\ref{I9}) is modified as
\begin{align}
\label{I99mim}
\frac{1}{\kappa^2} \left( R_{\mu\nu} - \frac{1}{2} g_{\mu\nu} R \right) = g_{\mu\nu}&\, \left[
 - \frac{1}{2}\, A (\phi,\chi) \partial_\rho \phi \partial^\rho \phi
 - B (\phi,\chi) \partial_\rho \phi \partial^\rho \chi
 - \frac{1}{2} \, C (\phi,\chi) \partial_\rho \chi \partial^\rho \chi - V (\phi,\chi)\right] \nn
&\, + A (\phi,\chi) \partial_\mu \phi \partial_\nu \phi
+ B (\phi,\chi) \left( \partial_\mu \phi \partial_\nu \chi
+ \partial_\nu \phi \partial_\mu \chi \right)
+ C (\phi,\chi) \partial_\mu \chi \partial_\nu \chi \nonumber \\
&\, + \frac{1}{2} g_{\mu\nu} \lambda \left\{ \left( 1 - \frac{b(\chi)}{\chi} \right) \partial_\rho \chi \partial^\rho \chi - 1 \right\}
 - \lambda \left( 1 - \frac{b(\chi)}{\chi} \right) \partial_\mu \chi \partial_\nu \chi \, ,
\end{align}
but we can always consider the solution with $\lambda=0$ and therefore the spacetime with the wormhole (\ref{1}) becomes an exact solution even if we add the term in (\ref{Smim}).

We may consider more general models where $\lambda$ does not vanish.
Applying the field equations (\ref{I99mim}) to the line element (\ref{1}), we obtain,
\begin{align}
\label{feBB}
\frac{b' \left( r \right) }{r^2} =&\, \frac{\left( \left( r\left( r-b \left( r \right) \right) C \left(r\right)
+2V \left(r\right) r^2 + b\left(r\right) \lambda\left(r\right) \left(2r-b\left(r\right) \right) \right) \e^{2\Phi \left( r \right)}
+A \left(r\right) r^2 \right) \kappa^2}{2 \e^{2\Phi \left( r \right)} r^2} \,,\nonumber\\
B \left(r\right)=&\, 0\,,\nonumber\\
\frac{2 r \left( r-b \left( r \right) \right) \Phi' \left( r \right) -b \left( r \right) }{r^3}
=&\,\frac{ \left( \left( r\left( r-b \left( r \right) \right) C \left( r \right) -2 V \left(r\right) r^2
 -\lambda\left(r\right) \left(2r^2-2rb\left(r\right)+b^2 \right) \right) \e^{2\Phi \left( r \right)}+A \left(r\right) r^2 \right) \kappa^2}{2 \e^{2\Phi \left( r \right)}r} \,,\nonumber\\
\frac{1}{r^3} & \left\{ 2 r^2 \left( r-b \left( r \right) \right) \Phi'' \left( r \right) +2 \left( r \left( r-b \left(r \right) \right)\Phi' \left( r \right)
 - \frac{b' \left( r \right)r}{2} + \frac{b \left( r \right)}{2} \right) \left( 1+ \Phi' \left( r \right) r \right) \right\} \nonumber\\
=&\, \frac{ \left( \left( r\left( b\left( r \right)-r \right) C \left(r\right) -2 V \left( r \right) r^2
+b\left(r\right)\lambda\left(r\right)(2r-b\left(r\right)) \right) \e^{2\Phi \left( r \right) } +A \left(r\right) r^2 \right) \kappa^2}
{\e^{2\Phi \left( r \right) } r} \, .
\end{align}
The solution of the above system takes the form:
\begin{align}
\label{sol1}
A \left(r\right) =&\, -\frac{\e^{\frac {r_0}{r}}}{4 \e^{\gamma r} \kappa^2 r^4} \left( 2 \e^{\gamma r_0}\gamma r^3 + 2 r_0 \e^{\gamma r_0}r
 - 4 r^2 \e^{\gamma r_0} - 2 r_0 r\e^{\gamma  r}+r_0\,\e^{\gamma r_0}\gamma r^2 + {r_0}^2\e^{ \gamma r_0}-{r_0}^2\e^{\gamma r} \right) \,, \nonumber\\
B \left(r\right)=&\, 0\,,\nonumber\\
C \left( r \right) =&\, \frac { \left( \e^{\gamma r}-\e^{\gamma r_0} \right) \lambda \left( r \right) }{\e^{\gamma r}}
 - \frac {2 \e^{\gamma r_0}\gamma r^3 - 6 r_0 \e^{ \gamma r_0} r + 4 r^2\e^{\gamma r_0} + 6 r_0 r\e^ {\gamma r} - r_0 \e^{\gamma r_0}\gamma r^2
  - {r_0}^2\e^{\gamma r_0}+{r_0}^2\e^{\gamma r}}{4{\kappa}^2 r^4 \left( \e^{\gamma r}-\e^{\gamma r_0} \right) }\,,\nonumber\\
V \left( r \right) =&\, -\frac{\lambda \left( r \right)}{2} -\frac{r_0 \e^{\gamma r_0}-2 \e^{\gamma r_0}r-r_0 \e^{\gamma r}+\e^{\gamma r_0}{r}^2\gamma}{2\e^{\gamma r}{\kappa}^2{r}^{3}}\, .
\end{align}
where $\lambda$ can take any value.
Of course, the solution (\ref{sol1}) gives the energy density and pressure components with those given by Eq.~(\ref{comp}) because the geometry is not changed.
By putting $r=\chi$ in (\ref{sol1}), we obtain a class of models where the wormhole spacetime given by (\ref{1}) with (\ref{WH}) and (\ref{TSBH1}) is an exact solution of the models.
Because $\lambda(\chi)=\lambda(r=\chi)$ is an arbitrary function, we may choose $\lambda(r)$ so that $C(\chi)=0$, that is
\begin{align}
\label{sol1C}
\lambda \left(r\right) = \frac {2 \e^{\gamma r_0}\gamma r^3 - 6 r_0 \e^{ \gamma r_0} r + 4 r^2\e^{\gamma r_0} + 6 r_0 r\e^ {\gamma r} - r_0 \e^{\gamma r_0}\gamma r^2
 - {r_0}^2\e^{\gamma r_0}+{r_0}^2\e^{\gamma r}}{4{\kappa}^2 r^4 \left( \e^{\gamma r}-\e^{\gamma r_0} \right)^2 \e^{-\gamma r} }\,,
 \end{align}
or so that $V(\chi)=0$,
\begin{align}
\label{sol1V}
\lambda \left( r \right) =\frac{r_0 \e^{\gamma r_0}-2 \e^{\gamma r_0}r-r_0 \e^{\gamma r}+\e^{\gamma r_0}{r}^2\gamma}{\e^{\gamma r}{\kappa}^2{r}^{3}}\, .
\end{align}
{ We should note that the ghost can be eliminated regardless of the choice (\ref{sol1C}) or (\ref{sol1V}) as confirmed in the next section}
and therefore the wormhole geometry in the model could be stable.

\section{Absence of ghost}\label{Sec4}

In order to investigate the (in)stability of the wormhole geometry, we focus on the model where $\lambda=0$ is a solution.
We now consider the perturbation around the solution (\ref{1}) with (\ref{WH}) and (\ref{TSBH1}) as follows,
\begin{align}
\label{pert1}
g_{\mu\nu} \to g_{\mu\nu} + h_{\mu\nu}\, , \quad
\phi\to \phi + \tau\, , \quad
\chi\to \chi + \xi\, , \quad
\lambda \to \lambda+\zeta\, .
\end{align}
Then Eq.~(\ref{I99mim}) in the background where $A = A(\chi)$, $B = 0$, $C = C(\chi)$, $V = V (\chi)$, and $\lambda=0$ gives,
\begin{align}
\label{pert2}
\frac{1}{\kappa^2} &\, \left[  \frac{1}{2} \left\{ \nabla_\mu \nabla^\rho h_{\nu\rho} + \nabla_\nu \nabla^\rho h_{\mu\rho} - \nabla^2 h_{\mu\nu}
 - \nabla_\mu \nabla_\nu \left( g^{\rho\sigma} h_{\rho\sigma} \right) - 2 R^{\sigma\ \rho}_{\ \nu\ \mu} h_{\sigma\rho}
+ R^\rho_{\ \mu} h_{\rho\nu} + R^\rho_{\ \nu} h_{\rho\mu} \right\} \right. \nonumber \\
&\, \left. - \frac{1}{2} h_{\mu\nu} R - \frac{1}{2} g_{\mu\nu} \left\{ - h_{\rho\sigma} R^{\rho\sigma} + \nabla^\rho \nabla^\sigma h_{\rho\sigma}
 - \nabla^2 \left( g^{\rho\sigma} h_{\rho\sigma} \right) \right\} \right] \nonumber \\
 =&\, h_{\mu\nu} \left[ - \frac{1}{2} A(\chi) \partial_\rho \phi \partial^\rho \phi - \frac{1}{2} C(\chi) \partial_\rho \chi \partial^\rho \chi - V(\chi) \right]
  - g_{\mu\nu} \left[ - \frac{1}{2} A(\chi) \partial^\rho \phi \partial^\sigma \phi - \frac{1}{2} C(\chi) \partial^\rho \chi \partial^\sigma \chi \right] h_{\rho\sigma} \nonumber \\
&\, - g_{\mu\nu} A(\chi) \partial_\rho \phi \partial^\rho \tau + A(\chi) \left( \partial_\mu \tau \partial_\nu \phi + \partial_\mu \phi \partial_\nu \tau \right)
 - g_{\mu\nu} C(\chi) \partial_\rho \chi \partial^\rho \xi + C(\chi) \left( \partial_\mu \xi \partial_\nu \chi + \partial_\mu \chi \partial_\nu \xi \right) \nonumber \\
&\, + \left[ g_{\mu\nu} \left\{ - \frac{1}{2} A'(\chi) \partial_\rho \phi \partial^\rho \phi - \frac{1}{2} C'(\chi) \partial_\rho \chi \partial^\rho \chi - V'(\chi) \right\}
+ A'(\chi) \partial_\mu \phi \partial_\nu \phi + C'(\chi) \partial_\mu \chi \partial_\nu \chi \right] \xi \nonumber \\
&\, - \zeta \left( 1 - \frac{b(\chi)}{\chi} \right) \partial_\mu \chi \partial_\nu \chi \, .
\end{align}
Here we also used the constraint (\ref{mimeticchi}).
Under the perturbation (\ref{pert1}), the constraint (\ref{mimeticchi}) has the following form,
\begin{align}
\label{pert3}
0= \left( 1 - \frac{b(\chi)}{\chi} \right) g^{\mu\nu} \partial_\mu \chi \partial_\nu \xi\, .
\end{align}
By using the background solution (\ref{1}) with (\ref{WH}) and (\ref{TSBH1}), the constraint (\ref{pert3}) gives,
\begin{align}
\label{pert4}
0 = \partial_r \xi \, ,
\end{align}
whose solution is $\xi=\xi(t,\theta,\phi)$ and $\xi$ does not depend on $r$.
Therefore if we put the boundary condition that $\xi\to 0$ when $r\to\infty$, we find $\xi$ identically vanishes,
\begin{align}
\label{pert5}
\xi=0\, .
\end{align}
This is because $\chi$ is not dynamical due to the mimetic constraint (\ref{mimeticchi}).
We now choose a condition to fix the gauge as follows,
\begin{align}
\label{pert6}
0 = \nabla^\mu h_{\mu\nu}\, .
\end{align}
Then Eq. (\ref{pert2}) with (\ref{pert5}) has the following form,
\begin{align}
\label{pert7}
\frac{1}{\kappa^2} &\, \left[  \frac{1}{2} \left\{ - \nabla^2 h_{\mu\nu}
 - \nabla_\mu \nabla_\nu \left( g^{\rho\sigma} h_{\rho\sigma} \right) - 2 R^{\sigma\ \rho}_{\ \nu\ \mu} h_{\sigma\rho}
+ R^\rho_{\ \mu} h_{\rho\nu} + R^\rho_{\ \nu} h_{\rho\mu} \right\}
 - \frac{1}{2} h_{\mu\nu} R - \frac{1}{2} g_{\mu\nu} \left\{ - h_{\rho\sigma} R^{\rho\sigma}
 - \nabla^2 \left( g^{\rho\sigma} h_{\rho\sigma} \right) \right\} \right] \nonumber \\
 =&\, h_{\mu\nu} \left[ - \frac{1}{2} A(\chi) \partial_\rho \phi \partial^\rho \phi - \frac{1}{2} C(\chi) \partial_\rho \chi \partial^\rho \chi - V(\chi) \right]
  - g_{\mu\nu} \left[ - \frac{1}{2} A(\chi) \partial^\rho \phi \partial^\sigma \phi - \frac{1}{2} C(\chi) \partial^\rho \chi \partial^\sigma \chi \right] h_{\rho\sigma} \nonumber \\
&\, - g_{\mu\nu} A(\chi) \partial_\rho \phi \partial^\rho \tau + A(\chi) \left( \partial_\mu \tau \partial_\nu \phi + \partial_\mu \phi \partial_\nu \tau \right)
 - \zeta \left( 1 - \frac{b(\chi)}{\chi} \right) \partial_\mu \chi \partial_\nu \chi \, .
\end{align}
By multiplying Eq. (\ref{pert7}) with $g^{\mu\nu}$ and using the mimetic constraint (\ref{mimeticchi}), we obtain
\begin{align}
\label{pert8}
\zeta=&\, - \frac{1}{\kappa^2} \left[ \nabla^2 \left( g^{\mu\nu} h_{\mu\nu} \right)
 - \frac{1}{2} \left( g^{\mu\nu} h_{\mu\nu} \right) R + 2 h_{\mu\nu} R^{\mu\nu} \right] \nonumber \\
&\, + \left( g^{\mu\nu} h_{\mu\nu} \right) \left[ - \frac{1}{2} A(\chi) \partial_\rho \phi \partial^\rho \phi - \frac{1}{2} C(\chi) \partial_\rho \chi \partial^\rho \chi - V(\chi) \right]
  - 4 \left[ - \frac{1}{2} A(\chi) \partial^\rho \phi \partial^\sigma \phi - \frac{1}{2} C(\chi) \partial^\rho \chi \partial^\sigma \chi \right] h_{\rho\sigma} \nonumber \\
&\, - 2 A(\chi) \partial_\rho \phi \partial^\rho \tau \, .
\end{align}
This tells that $\zeta$ is not an independently propagating mode.

We may consider the perturbation of (\ref{I10})
\begin{align}
\label{I10B}
0 = A(\chi) \nabla^2 \tau - A(\chi) h_{\mu\nu} \nabla^\mu \nabla^\nu \phi
 - \frac{1}{2} A(\chi) \nabla^\rho \left( g^{\mu\nu} h_{\mu\nu} \right) \partial_\rho \phi
 + A_\chi \partial_\mu \tau \partial^\mu \chi \, .
\end{align}
Therefore $\tau$ behaves as a massless mode.
And as long as the condition (\ref{N3}) is satisfied, which tells $A$ is positive, the scalar mode $\tau$ is not a ghost but a canonical scalar.

The remaining mode could be only massless spin-two mode corresponding to the standard graviton.
Therefore there is no mode generating ghost instability.
{ This result is valid whether $\lambda=0$ or $\lambda\neq 0$ because the constraint (\ref{mimeticchi}) or (\ref{pert3}) is always obtained regardless
of the background value of $\lambda$.
}

In order to investigate the causality, we often use the speed of sound in a fluid.
The radial and tangential speeds of sound could be defined by
\begin{align}
\label{eq:sound_speed}
v_r^2 = \frac{dp_r}{d \rho}= \frac{p'_r}{\rho'}\, , \quad
v_t^2 = \frac{dp_t}{d\rho}= \frac{p'_t}{\rho'}\,.
\end{align}
As we have seen, however, nothing corresponds to the sound wave.
The existing waves are gravitational waves and massless scalar waves, whose propagating speeds are different from the above sound speeds.
Therefore the arguments of the sound speed are not applicable here.
In other words, the dynamics of the fields cannot be approximated by the dynamics of the fluid, in general.
Even if we can approximate the field(s) by any fluid the EoS could not be so simple.
In the simple case, the pressure $p$ only depends on the energy density $\rho$ but in general, $p$ depends on other parameters, and therefore the expressions in
(\ref{eq:sound_speed}) are not approved for the general cases.
Even in our model, because $p_\mathrm{r}\neq p_\mathrm{t}$, the pressure depends on the direction and therefore the field cannot be expressed by the perfect fluid,
which has no direction dependence.

\section{Summary and discussions}\label{Sec5}

Studying stable wormholes in the context of Einstein's theory of GR  with two scalar fields is a complex and challenging endeavor.
Such research would likely involve a deep dive into theoretical physics, differential geometry, and advanced mathematical techniques.

Stable wormholes have been a popular subject in science fiction, often depicted as portals to other parts of the Universe or alternate dimensions.
While these portrayals are speculative, the idea of traversable wormholes has captured the imagination of both scientists and the general public.
Investigating the theoretical possibility of stable wormholes can be seen as a step toward understanding the Universe's potential intricacies.

In this study, we constructed the model where the standard and traversable wormhole geometries are included in the exact solutions.
We have used the formulation in \cite{Nojiri:2020blr}, where it has been shown
how the model reproducing general spherically symmetric and even time-dependent solutions can be constructed.
The model based on the original formulation in \cite{Nojiri:2020blr}, however, includes ghosts.

In order to eliminate the ghosts, we have imposed the mimetic constraint on the scalar field so that the ghost fields become non-dynamical.
As a result, although the energy conditions are broken, we have obtained models without the instability due to the ghosts.
This could tell that there are stable models even if the energy conditions are broken.

In this study, we succeeded in constructing realistic a stable wormhole using Einstein GR with two scaler fields.
Can this procedure be applied in the frame of $f(R)$ gravitational theory with two scalar fields or in the frame of Gauss-Bonnet theory with two scalar fields?
All these questions may be answered elsewhere.
%More general arguments could be given in future works.

%%%%%%%%%%%%%%%%%%%%%%%%%%%%%%%%%%%%%%%%%%%%%%%%%%%%%%%%%%%%%%%%%%%%%%%%%%%%%%%%%%%%%%
%\bibliographystyle{apsrev}
%\bibliography{JRPHSRef}
%%%%%%%%%%%%%%%%%%%%%%%%%%%%%%%%%%%%%%%%%%%%%%%%%%%%%%%%%%%%%%%%%%%%%%%%%%%%%%%%%%%%%%

%%%%%%%%%%%%%%%%%%%%%%%%%%%%%%%%%%%%%%%%%%%%%%%%%%%%%

\end{document}